# Mapping the Authorized Boundary: A Comparative Policy-Vignette Study of Generative AI Governance in Australian Higher Education

Biranchi Poudyal

Faculty of Arts and Society, Charles Darwin University, Australia

ORCID: https://orcid.org/0000-0002-7210-5480

## Abstract

Australian universities regulate students 'use of generative artificial intelligence (GenAI) through overlapping policies, procedures, guidance, and assessment instructions, but these environments may classify identical conduct differently. We applied 15 standardized student-use vignettes to the public policy environments of 20 Australian universities and produced 300 university-case classifications. We analyzed binding instruments (Layer A) separately from the full official environment (Layer B), then classified each combination as clearly permitted, permitted with conditions, potential policy breach, clearly prohibited, or indeterminate. We measured cross-university divergence with normalized Shannon entropy. We classified 120 combinations (40.0%) as clearly prohibited, 97 (32.3%) as potential policy breaches, 27 (9.0%) as permitted with conditions, and 56 (18.7%) as indeterminate; none met the strict threshold for clearly permitted. Disclosed language rewriting and a disclosed AI-drafted paragraph produced the highest divergence, whereas an explicit assessment prohibition produced unanimity. Binding instruments remained silent on GenAI in 100 combinations, and guidance clarified 88. The primary researcher led the coding with AI assistance and manually reviewed all 201 queued rows. An independent human second coder, who used no AI assistance, coded a blind 75-row sample. Overall agreement reached 57.3% (unweighted Cohen's kappa = 0.395; bootstrap 95% CI [0.244, 0.539]). The findings distinguish permission-gated from disclosure-based architectures and show that written policies regulate the retention of AI-generated text more clearly than process-only assistance does. Comparative policy-vignette testing evaluates whether written governance supports defensible classifications; it does not predict misconduct or enforcement decisions.

## 1. Introduction

Generative AI has quickly become part of how students plan, find information, draft, revise, and check academic work. Universities have responded with academic-integrity policies, misconduct procedures, community-wide GenAI guidance, staff resources, student webpages, and assessment instructions. International studies show substantial variation: institutions encourage use, permit it under conditions, prohibit it in selected contexts, or delegate decisions to educators (An et al., 2025; Jin et al., 2025; McDonald et al., 2025; Parker et al., 2025).

A formal policy does not by itself draw a clear boundary between assistance and misconduct. A rule against unauthorized assistance may not explain how it applies to an AI-generated outline, language rewriting, or source discovery. Guidance may encourage experimentation but direct students to assessment instructions. A disclosure rule may explain acknowledgment without granting permission. Older rules on fabrication, falsification, or contract cheating may still govern AI-enabled conduct even when they never mention GenAI. Students and educators, therefore, navigate a policy environment rather than one document. If institutions deliberately choose different educational models, different classifications may reflect legitimate policy diversity. If ambiguity, conflicting guidance, or missing defaults cause the difference, the written system lacks operational clarity. Comparative policy vignettes hold constant conduct and expose that distinction.

Australia offers a useful setting because its universities have mature academic-integrity systems and national regulators now emphasize assessment reform, assurance of learning, and critical judgment in GenAI-enabled settings (Bretag et al., 2011; Lodge et al., 2023, 2026; Tertiary Education Quality and Standards Agency [TEQSA], 2025). Recent research, however, reports substantial differences in institutional policy details, delegated authority, and classroom implementation (De Maio, 2024; Jiang et al., 2025; Parker et al., 2025).

### 1.1 International relevance

Although Australia offers a defined policy setting, it does not face a unique governance problem. Cross-national research shows that universities combine institutional policy, central guidance, and educator-specific rules with varying permissions, disclosures, and discretion (Jin et al., 2025; McDonald et al., 2025; Parker et al., 2025). The analytical questions in this study, therefore, travel across systems: where does authority sit, what default applies when local instructions remain silent, and can a reasonable reader identify a defensible classification?

### 1.2 Study objectives and contributions

We applied 15 standardized vignettes to the public policy environments of 20 Australian universities. The vignettes varied in AI contribution, disclosure, false declarations, verification, and explicit task prohibition. Holding student conduct constant allowed us to isolate how written institutional environments shaped classification.

> RQ1. To what extent do Australian university policy environments yield consistent, divergent, or indeterminate classifications when applied to the same standardized GenAI student-use cases?
>
> RQ2. How do policy authority, permission requirements, disclosure rules, authorship, deception, verification, and educator discretion explain differences in institutional classifications?

This study makes three contributions. It maps where written policy environments converge, diverge, or remain ambiguous while student conduct stays constant. It develops comparative policy-vignette testing as a method for evaluating written governance. It also proposes a six-question framework that tests task rules, defaults and authority, disclosure, contribution and authorship, verification, and policy hierarchy. The study evaluates what public written governance supports, not what an institution would decide after considering evidence, intent, procedure, and local context.

## 2. Literature review

## 2.1 What policy catalogs reveal

Early university GenAI research documented a changing mix of prohibition, cautious acceptance, and integration. McDonald et al. (2025) found that many of the 116 United States research universities encouraged the use of GenAI and provided classroom resources, while leaving implementation to educators. Jin et al. (2025) identified similar efforts across 40 universities in six world regions to combine academic-integrity safeguards, ethical use, training, and pedagogical integration. Parker et al. (2025) examined 343 universities in five countries and found widespread instructor discretion alongside regional differences in rationale and detail.

These findings resist a simple permit-ban scale. A university may promote AI literacy yet prohibit AI in one assessment; it may require prior approval or permit use by default with acknowledgment. Institutional materials also address different audiences and purposes, including teaching, research, administration, and student services (An et al., 2025). This distribution creates a rule system in which authority and purpose shape the answer.

Australian policy analyses provide useful benchmarks. Bretag et al. (2011) identified access, approach, responsibility, detail, and support as core elements of an exemplary academic integrity policy. De Maio (2024) extended that framework to GenAI and found differences in currency, flexibility, and clarity across four institutions. Perkins and Roe (2024), analyzing 142 policies, found that existing integrity language rarely addressed GenAI and recommended explicit technological references. These studies show which policies are included, but they do not test what readers can conclude when institutions apply those policies to the same contested conduct.

### 2.2 Why policy catalogs do not solve the classification problem

GenAI can assist at several points in academic work without leaving the same amount or type of text: ideation, outlining, source discovery, translation, editing, sentence rewriting, paragraph generation, or major drafting. Students also judge these uses differently; Johnston et al. (2024), for example, found greater acceptance of language support than whole-essay generation. Classification, therefore, depends on what the tool did, what the student retained, and what the student did before and after using it.

Recent integrity scholarship shifts attention from detection to accountability and visible judgment. Eaton (2023) describes a post-plagiarism context in which technology is part of everyday knowledge creation and entails renewed ethical obligations. Sharma (2026) argues that GenAI-enabled assessment should make learners' evaluative judgments visible while holding them responsible for verification and interpretation. Australian regulatory guidance likewise emphasizes evidence of learning and assessment designs that allow designated AI use (Lodge et al., 2023; TEQSA, 2025).

Disclosure matters, but it serves several purposes: it can acknowledge assistance, create an audit trail, or satisfy a precondition for use. It does not necessarily grant permission. A student can disclose prohibited conduct, and a policy can permit conduct while still requiring a declaration. Qu et al. (2025) show that students may strategically conceal use or disagree about when AI assistance constitutes authorship. Treating disclosure as authorization obscures the prior question of permission.

False declarations add deception, but their effect depends on the underlying rule. If a policy already clearly prohibits substantial AI drafting, a false declaration aggravates the conduct without changing its primary category. If the drafting rule remains ambiguous, a misrepresentation rule may supply a determinate answer. Policy catalogs cannot reveal these case-level effects.

## 2.3 Operational governance: layers, defaults, and discretion

University governance operates through a hierarchy. Policies, rules, procedures, and ordinances generally carry more authority than webpages or teaching guides, while guidance often supplies details that formal instruments omit. Course and assessment instructions can control a specific task. Regulatory intermediacy theory treats governance as a relationship among rule makers, intermediaries, and regulated targets (Abbott et al., 2017): institutional authorities set general rules, organizational units and educators translate them, and students act within the resulting system. General rules still require interpretation and may prove too broad or narrow (Black, 1997). Problems arise at the handoffs. Policy delegates to educators, guidance tells students to check local instructions, and those instructions may remain silent or conflict. A direction to 'ask your educator' provides no default when the educator has not spoken; a rule that allows AI only with authorization does. Students often recognize classroom rules more readily than institutional policies, while educators report constraints in training and implementation (Alsharefeen & Al Sayari, 2025; Jiang et al., 2025). Good governance need not remove discretion, but it should define its scope, identify the decision-maker, and specify a fallback. This distinction also guards against two common analytical errors. Researchers should not treat guidance as binding merely because it provides a clear answer, nor should they treat one policy's silence as proof that the institution lacks a rule. A complete classification requires a documented route through authority, guidance, and task-level material.

### 2.4 Why operational governance needs testing

Policy studies describe institutional content, tone, and distribution, but they rarely test whether the same conduct receives the same classification. Standardized vignettes can act as diagnostic probes because they hold core facts constant while varying selected attributes (Hughes & Huby, 2002; Wallander, 2012). This study applies vignettes to documents rather than respondents. Divergence measures how universities distribute classifications across cases. Indeterminacy records when the collected documents do not support a defensible answer. Universities may converge on indeterminacy, or they may diverge among several clear answers. Keeping the concepts separate prevents readers from confusing cross-institutional disagreement with a lack of written guidance.

### 2.5 A three-dimensional framework for policy quality

Operational governance requires three linked dimensions. Policy layers locate the controlling authority. Default rules preserve continuity when a lower-level actor or document remains silent. Operational clarity asks whether a reasonable reader can reach a defensible classification after navigating those layers and defaults. No dimension can replace another: authority without a default leaves silence unresolved; a default without authority may not apply; clear wording in one document may not create an integrated system. The coding framework operationalizes these dimensions, and the Discussion develops them into a model of how written governance shapes the conditions for decision-making. The framework treats discretion as a designed feature rather than a defect. Discretion becomes a clarity problem only when written materials leave their holder, scope, communication route, or fallback unspecified. This focus permits substantive differences among institutions while still asking whether each system coherently communicates its choice.

# 3. Methods

## 3.1 Design and unit of analysis

We used a fully crossed comparative policy-vignette design. The unit of analysis was the university-case combination. We applied 15 fixed vignettes to 20 universities, producing 300 combinations, and assigned each combination to one of five outcome categories. Public institutional documents served as evidence of the written policy environment, not of institutional practice or enforcement.

Cases 1-12 crossed four levels of AI contribution (outline, language rewriting, one retained AI-drafted paragraph, and a major AI-generated first draft) with three disclosure conditions (full disclosure, no disclosure, and a false declaration). Cases 13-15 acted as targeted stress tests rather than a complete factorial sequence. Case 13 examined AI-assisted source discovery followed by independent verification; Case 14 examined fabricated AI-generated references without verification; and Case 15 repeated disclosed outline use under an explicit assessment prohibition.

### 3.2 Institutional sample and document corpus

We built a preliminary pool of 30 universities to maximize variation across state or territory, metropolitan or regional setting, Group of Eight status, and public or private status, then purposively selected 20 while retaining that variation. The sample does not represent all Australian providers. It comprised Adelaide University, Australian National University, Bond University, Charles Darwin University, Charles Sturt University, Curtin University, Deakin University, James Cook University, Macquarie University, Monash University, Queensland University of Technology, RMIT University, UNSW Sydney, the universities of Melbourne, Newcastle, Queensland, Sydney, Tasmania, Technology Sydney, and Western Australia. We treated Adelaide University as the successor institution after its January 1, 2026, merger. We collected public documents on July 21-22, 2026. The index records 150 documents: 138 included documents (59 binding instruments and 79 guidance documents) and 12 archived or superseded items retained only for provenance. Each institution contributed between two and fourteen included documents.

### 3.3 Policy layers and classification framework

We coded two layers. Layer A comprised binding policies, procedures, statutes, rules, and formal misconduct provisions. Layer B comprised the full official environment: Layer A plus official guidance for students, staff, faculty, and assessment. We analyzed Layer A first, then reviewed Layer B for clarification or conflict. We based the final classification on the full official environment while recording the layer that supplied the operative rule. We did not treat silence in a single binding document as institutional silence until we had reviewed all relevant Layer A instruments. The preregistered categories were: Clearly permitted (direct, unconditional authorization); Permitted with conditions (authorization subject to stated requirements); Potential policy breach (a relevant provision may apply, but its scope or effect remains ambiguous); Clearly prohibited (a provision directly prohibits the conduct); and Indeterminate (the collected official material remains insufficient or contradictory). The categories answer different questions about sanctions or findings of misconduct. Potential policy breach records a plausible but uncertain application; Indeterminate records the absence of a supported answer. This separation prevents coders from forcing ambiguous documents into a determinate category and preserves uncertainty, so that policy designers can examine.

The codebook set a deliberately strict threshold for clearly permitted. General encouragement did not qualify when educator approval, disclosure, task instructions, or another condition still applied. We coded disclosure separately from substantive authorization. Additional fields recorded prior permission, educator discretion, authorship, deception, verification, policy conflict, clarity, confidence, supporting clauses, and document authority-every determinative classification required page-based support.

### 3.4 Pilot, coding, and verification

A three-university pilot exposed differences between systems with explicit defaults and systems that delegated decisions without a fallback. We clarified four rules before full coding: scope ambiguity could support Indeterminate; coders should record policy conflict separately; general fabrication provisions applied to Case 14; and coders had to verify the page for each determinative quotation. Guidance could clarify a silent binding layer, but could not override a specific binding rule.

The researcher led the primary coding with assistance from a Claude Code AI agent. The tool helped retrieve candidate evidence and organize preliminary classifications, rationales, document titles, pages, and quotations; the researcher retained interpretive responsibility for every final code. A rule-based queue identified 201 rows for human review due to ambiguity, conflict, reliance on guidance, or other verification concerns. The researcher personally reviewed all 201 rows on July 22, 2026, before freezing the dataset. The final dataset separately marks 184 rows as requiring manual verification under a narrower row-level field; that field and the broader review queue measure different things. Because the workflow did not retain a row-level change log, the study does not report how many classifications changed during review. The primary workflow, therefore, combined computational assistance with mandatory human judgment at every contested point. The researcher checked cited documents, page references, quotations, authority levels, and alternative interpretations before accepting a queued row. The frozen dataset records the researcher's conclusions rather than the model's autonomous outputs.

An independent human second coder then coded a stratified sample of 75 rows (25% of the dataset), covering all 20 universities and all 15 cases. The second coder used no AI assistance. Before coding, the second coder could not see the primary overall or disclosure classifications, rationales, or confidence ratings. This audit, therefore, measured intercoder reliability between the human-led, AI-assisted primary coding and independent human coding.

We calculated exact agreement and unweighted Cohen's kappa for the nominal five-category overall outcome. We did not use weighted kappa because Indeterminate represents a missing supported answer rather than a severity level. We estimated a percentile 95% confidence interval from 10,000 row-level bootstrap resamples with a fixed seed and analyzed disclosure classifications separately. After revealing the primary labels, we documented dispositions for all 32 disagreements but did not alter the frozen 300-row dataset. We used exact agreement because every category change matters under a nominal scheme. The fixed bootstrap seed makes the confidence interval reproducible. We retained both coders' original labels so readers can inspect the confusion matrix and assess whether the main claims depend on contested boundaries.

Structural checks confirmed a complete 20-by-15 grid with no duplicate or missing combinations and only permitted category values. Every one of the 300 rows contained a supporting clause, and all determinate rows contained evidence. The document audit linked 300 coded rows to supporting clauses and found no page-reference gaps. We also reconciled the frozen dataset against every reported table and result. We also confirmed that the public document index contained 150 unique entries, that the included corpus contained 138 documents, and that the evidence audit recorded zero page-reference gaps. Re-running the supplied scripts reproduced the descriptive, divergence, matched-case, policy-layer, evidence-log, theme, and agreement outputs.

### 3.5 Analysis

We reported frequencies and percentages for classifications, disclosure outcomes, authority variables, and institutional ambiguity. For each case, we measured cross-university divergence with normalized Shannon entropy:

$$\mathbf{H = -\Sigma\ p_i \ln(p_i) / \ln(5),}$$

where p_i denotes the proportion of universities in classification i across five categories, the index ranges from 0 (one category) to 1 (an even distribution across all five). It measures dispersion, not policy quality or ambiguity.

For matched comparisons, we ordered the four determinate categories from Clearly permitted to Clearly prohibited and kept Indeterminate separate. We coded each paired change as unchanged, more restrictive, more permissive, newly indeterminate, resolved from indeterminate, or indeterminate in both cases.

We examined reasoning and supporting clauses across all 300 rows and checked qualitative themes against structural variables and quantitative results. Theme-support counts identify overlapping evidence rows and do not form mutually exclusive prevalence estimates.

We ran the analysis in Python 3.12.3. Matplotlib 3.9.2 and NumPy 2.2.6 generated the figures, and python-docx 1.1.2 supported document processing. Online Resource 1 provides the instrument and coding protocol, Online Resource 2 provides the data and audit workbook, and Online Resource 3 provides the analysis scripts and machine-readable inputs.

## 3.6 Scope and ethics

The study used public institutional documents and involved no human participants or personal data. It evaluates the operational clarity of written governance, not practice or enforcement, and does not predict misconduct decisions. Public documents may change after collection, and internal materials may yield different answers. The purposive sample and fixed cases support analytic comparison but not sector-wide statistical generalization.

# 4. Results

## Key findings

Four findings anchor the analysis. No combination met the strict threshold for Clearly permitted. Disclosed language rewriting and a disclosed AI-drafted paragraph produced the greatest divergence. Guidance resolved 88 of the 100 combinations in which binding instruments remained silent on GenAI. Finally, written policies more clearly regulated the retention of AI-generated text than process-only assistance, while an explicit task prohibition produced unanimity.

## 4.1 Overall classification pattern

The written policy environments produced mostly restrictive classifications. Clearly prohibited accounted for 120 of 300 combinations (40.0%), Potential policy breach for 97 (32.3%), Permitted with conditions for 27 (9.0%), and Indeterminate for 56 (18.7%). No combination met the strict threshold for Clearly permitted (Table 1).

**Table 1. Overall policy-vignette classifications**

| Classification | n | % |
|---|---|---|
| Clearly permitted | 0 | 0.0 |
| Permitted with conditions | 27 | 9.0 |
| Potential policy breach | 97 | 32.3 |
| Clearly prohibited | 120 | 40.0 |
| Indeterminate | 56 | 18.7 |
| Total | 300 | 100.0 |

Note. Percentages use all 300 university-case combinations. ‘Clearly permitted’required direct, unconditional authorization.

These results do not show that the sampled universities uniformly oppose GenAI. Many institutional materials encourage AI learning or permit use after educator approval, task authorization, or acknowledgment. The codebook classified those cases as conditional rather than unconditional permission.

Disclosure did not map directly onto substantive outcomes. The disclosure field recorded 117 satisfied requirements, 80 false declarations, 64 required-but-absent disclosures, 27 unclear positions, and 12 cases in which policy did not require disclosure. Permission, authorship, and task instructions could still produce different substantive classifications for the same disclosure status.

## 4.2 Second-coder audit

The independent human second coder agreed with 43 of 75 primary overall classifications (57.3%). Unweighted Cohen’s kappa reached 0.395 (bootstrap 95% CI [0.244, 0.539]), which indicates moderate agreement on the demanding five-category nominal scale. Disclosure agreement reached 88.0% (kappa = 0.824). These pre-adjudication results support a sensitivity audit, not a claim of highly stable classification.

The 32 overall disagreements clustered at two boundaries. Eleven compared primary Potential policy breach with second-coder Clearly prohibited, and ten compared primary Indeterminate with second-coder Permitted with conditions. The remaining eleven cases spanned five smaller transition types. Coders, therefore, struggled most with deciding when a rule clearly prohibited conduct and when affirmative authorization overcame indeterminacy. Specifically, five disagreements moved from primary Clearly prohibited to second-coder Potential policy breach; three moved from Potential policy breach to Indeterminate; and one each compared Indeterminate with Clearly prohibited, Permitted with conditions with Clearly prohibited, and Potential policy breach with Permitted with conditions. This distribution supports caution about labels at the edges of authorization and prohibition rather than about the dataset's basic structure.

We retained all disagreements in the audit and left the frozen primary dataset unchanged. Applying reconciled outcomes only to one quarter of the dataset would have mixed coding standards. The audit instead marks category boundaries that require cautious interpretation.

## 4.3 Divergence across cases

Case-level distributions varied substantially (Table 2; Figure 1). Cases 4 (disclosed language rewriting) and 7 (one disclosed AI-drafted paragraph) produced the highest divergence (H = 0.801). Each produced 8 Potential policy breaches, 5 permitted with conditions, 5 clearly prohibited, and 2 Indeterminate classifications. These moderate-contribution cases separated permission-gated from disclosure-based architectures.

**Table 2. Classification, distribution, and cross-university divergence by vignette**

| Case | Scenario | CP | PWC | PPB | CPr | Ind. | H |
|---|---|---|---|---|---|---|---|
| 1 | Outline; disclosed | 0 | 4 | 4 | 1 | 11 | 0.697 |
| 2 | Outline: not disclosed | 0 | 2 | 5 | 2 | 11 | 0.706 |
| 3 | Outline: false declaration | 0 | 1 | 4 | 7 | 8 | 0.749 |
| 4 | Language rewriting; disclosed | 0 | 5 | 8 | 5 | 2 | 0.801 |
| 5 | Language rewriting; not disclosed | 0 | 1 | 10 | 8 | 1 | 0.629 |
| 6 | Language rewriting; false declaration | 0 | 0 | 8 | 11 | 1 | 0.525 |
| 7 | AI paragraph; disclosed | 0 | 5 | 8 | 5 | 2 | 0.801 |
| 8 | AI paragraph; not disclosed | 0 | 1 | 10 | 8 | 1 | 0.629 |
| 9 | AI paragraph; false declaration | 0 | 0 | 9 | 10 | 1 | 0.532 |
| 10 | Major draft; disclosed | 0 | 4 | 7 | 8 | 1 | 0.749 |
| 11 | Major draft; not disclosed | 0 | 1 | 8 | 11 | 0 | 0.525 |
| 12 | Major draft; false declaration | 0 | 0 | 7 | 13 | 0 | 0.402 |
| 13 | Verified source discovery | 0 | 3 | 1 | 0 | 16 | 0.381 |
| 14 | Fabricated references | 0 | 0 | 8 | 11 | 1 | 0.525 |
| 15 | Explicit task prohibition | 0 | 0 | 0 | 20 | 0 | 0.000 |

Note. CP = Clearly permitted; PWC = Permitted with conditions; PPB = Potential policy breach; CPr = Clearly prohibited; Ind. = Indeterminate. H denotes normalized Shannon entropy across five categories.

**Fig. 1 Classification distribution by vignette**

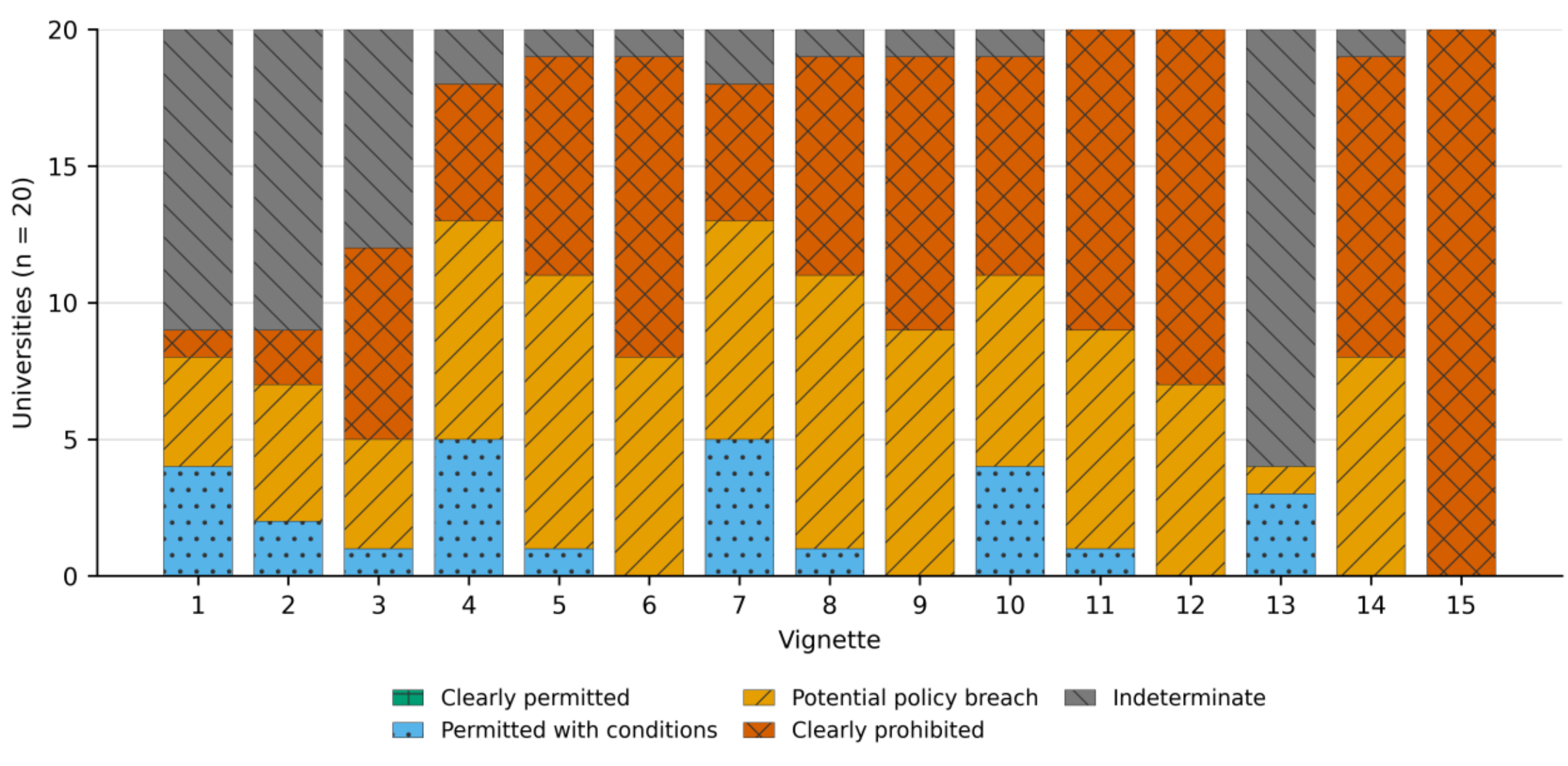

Note. Each bar contains 20 university classifications. No vignette produced a clearly permitted classification.

Case 15 produced complete agreement: all 20 universities classified the disclosure of outline generation as clearly prohibited when an assessment instruction explicitly banned the use of AI, so H = 0. Case 13 shows why entropy and ambiguity differ. Sixteen universities classified verified AI-assisted source discovery as Indeterminate, three as Permitted with conditions, and one as Potential policy breach. Its relatively low entropy (0.381) reflected convergence mainly on the absence of a supported answer.

Process-only cases produced more indeterminacy than cases with retained AI-generated language. Cases 1 and 2 (outline use with and without disclosure) each produced 11 Indeterminate classifications, and Case 13 produced 16. No major-drafting case produced more than one Indeterminate outcome, and Case 12 produced none.

The university-by-case heatmap shows a structured pattern (Figure 2). Some universities applied permission-gated defaults, others treated acknowledgment as a condition of otherwise permitted use, and others required missing course-level information. Every university recognized the explicit task rule in Case 15.

**Fig. 2 University-by-vignette classification heatmap**

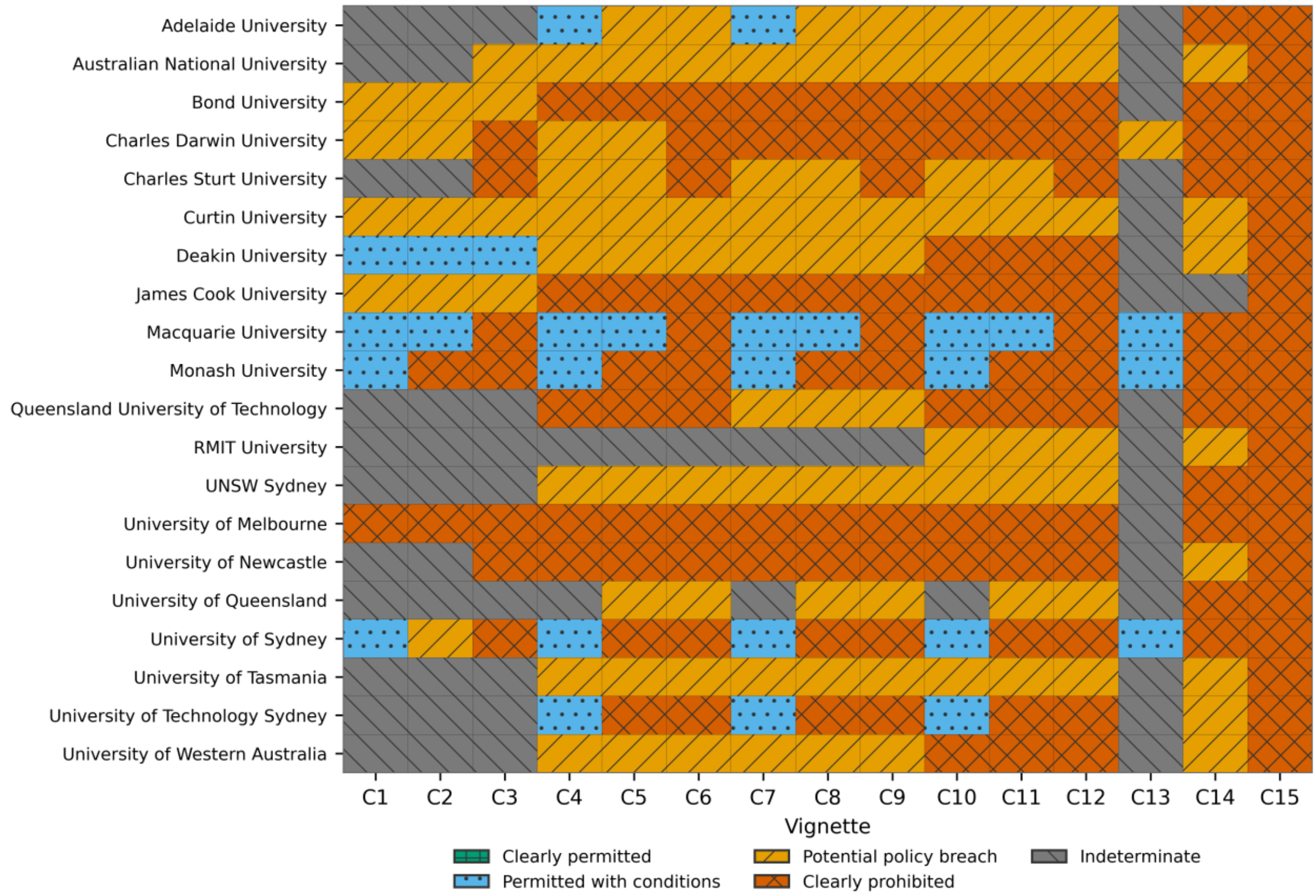

Note. Rows represent universities, and columns represent the 15 fixed vignettes. Category colors match Figure 1.

## 4.4 Matched case comparisons

Matched comparisons show that disclosure alone often changed little (Table 3). Removing disclosure left 14 of 20 classifications unchanged for rewriting and paragraph generation, and 16 unchanged for major drafting; two to four universities became more restrictive. In the outline pair, 11 universities remained Indeterminate in both cases.

**Table 3. Selected matched-case transitions across 20 universities**

| Comparison | Same | More restr. | More perm. | New ind. | Resolved | Both ind. |
|---|---|---|---|---|---|---|
| Full → no disclosure: outline (C1→C2) | 7 | 2 | 0 | 0 | 0 | 11 |
| Full → no disclosure: rewriting (C4→C5) | 14 | 4 | 0 | 0 | 1 | 1 |
| Full → no disclosure: paragraph (C7→C8) | 14 | 4 | 0 | 0 | 1 | 1 |
| Full → no disclosure: major draft (C10→C11) | 16 | 3 | 0 | 0 | 1 | 0 |
| Outline → rewriting: disclosed (C1→C4) | 6 | 3 | 0 | 0 | 9 | 2 |
| Paragraph → major draft: disclosed (C7→C10) | 14 | 4 | 0 | 0 | 1 | 1 |
| Verified sources → fabricated refs (C13→C14) | 0 | 4 | 0 | 0 | 15 | 1 |
| Disclosed outline → explicit ban (C1→C15) | 1 | 8 | 0 | 0 | 11 | 0 |

Note. The analysis keeps Indeterminate outside the determinate severity sequence. Online Resource 2 contains the complete matched-comparison table.

False declarations also changed fewer classifications than expected: 16 rewriting, 17 paragraph-generation, and 18 major-drafting comparisons remained unchanged. Deception still mattered, but many policies already prohibited the underlying AI contribution, so the primary category could not become more restrictive.

Changes in contribution mattered most when conduct moved from process support to retained text. Moving from outline generation to language rewriting resolved indeterminacy for nine universities with disclosure, ten without disclosure, and seven with a false declaration. Moving from rewriting to one AI-drafted paragraph usually left the classification unchanged, suggesting that many rules already treated retained machine-generated language as the relevant boundary.

Verification failure and explicit task authority produced the strongest shifts. Between verified source discovery (Case 13) and fabricated references (Case 14), 15 universities moved from Indeterminate to a determinate outcome, and four became more restrictive. Between disclosed outline use under silent instructions (Case 1) and an explicit prohibition (Case 15), 11 cases were resolved from Indeterminate, and eight became more restrictive.

## 4.5 Policy layers and institutional ambiguity

Layer A alone, or agreement between Layers A and B, resolved 188 of 300 combinations (62.7%); Layer B supplied the main operative rule in 109 (36.3%). Binding instruments remained silent on GenAI in 100 combinations, and guidance clarified 88. We recorded policy conflict in 24 combinations (8.0%), educator discretion in 204 (68.0%), and an express prior-permission requirement in 128 (42.7%). The remaining three combinations recorded direct conflict between Layers A and B. These counts show why a search for GenAI language in formal policy alone would miss many operative answers: guidance frequently completed the classification pathway, but its contribution varied across institutions and cases.

Ambiguity varied sharply by institution (Table 4; Figure 3). RMIT University produced 10 Indeterminate cases (66.7%), and the University of Queensland produced 7 (46.7%). Charles Darwin, Macquarie, Monash, and Sydney produced none. Adelaide, Bond, Charles Darwin, Monash, Queensland University of Technology, and Sydney resolved all 15 cases at Layer A.

**Table 4. Ambiguity and governance-layer indicators by the university**

| University | Indet. % | Binding silent % | Layer B % | Discretion % | Conflicts n |
|---|---|---|---|---|---|
| RMIT University | 66.7 | 100.0 | 93.3 | 86.7 | 0 |
| University of Queensland | 46.7 | 86.7 | 80.0 | 80.0 | 0 |
| University of Tasmania | 26.7 | 100.0 | 93.3 | 93.3 | 0 |
| UNSW Sydney | 26.7 | 0.0 | 26.7 | 86.7 | 3 |
| University of Technology Sydney | 26.7 | 0.0 | 26.7 | 46.7 | 6 |
| University of Western Australia | 26.7 | 13.3 | 26.7 | 86.7 | 3 |
| Adelaide University | 26.7 | 33.3 | 0.0 | 6.7 | 0 |
| Queensland University of Technology | 26.7 | 0.0 | 0.0 | 80.0 | 0 |
| Australian National University | 20.0 | 93.3 | 93.3 | 86.7 | 0 |
| Charles Sturt University | 20.0 | 60.0 | 66.7 | 86.7 | 4 |
| University of Newcastle | 20.0 | 0.0 | 26.7 | 86.7 | 4 |
| James Cook University | 13.3 | 6.7 | 6.7 | 86.7 | 3 |
| Curtin University | 6.7 | 100.0 | 93.3 | 86.7 | 0 |
| Deakin University | 6.7 | 6.7 | 20.0 | 0.0 | 0 |
| University of Melbourne | 6.7 | 0.0 | 6.7 | 86.7 | 1 |
| Bond University | 6.7 | 0.0 | 0.0 | 100.0 | 0 |
| Macquarie University | 0.0 | 66.7 | 66.7 | 0.0 | 0 |
| Charles Darwin University | 0.0 | 0.0 | 0.0 | 86.7 | 0 |
| Monash University | 0.0 | 0.0 | 0.0 | 86.7 | 0 |
| University of Sydney | 0.0 | 0.0 | 0.0 | 0.0 | 0 |

Note. Percentages are used for the 15 vignettes at each university. Layer B % records the share resolved primarily through guidance after Layer A remained silent. Conflict n records inter- or intra-layer conflicts.

**Fig. 3 Institutional ambiguity and reliance on official guidance**

RMIT University
University of Queensland
University of Tasmania
UNSW Sydney
University of Technology Sydney
University of Western Australia
Adelaide University
Queensland University of Technology
Australian National University
Charles Sturt University
University of Newcastle
James Cook University
Curtin University
Deakin University
University of Melbourne
Bond University
Macquarie University
Charles Darwin University
Monash University
University of Sydney
Indeterminate classifications
Decisions relying mainly on Layer B

Note. Percentages are used for the 15 vignettes at each university. Layer B reliance means guidance supplied the primary answer after the binding layer remained silent.

Binding silence did not determine ambiguity. Curtin, Tasmania, and RMIT had silent binding layers across all 15 cases, yet their ambiguity rates were 6.7%, 26.7%, and 66.7%, respectively. Curtin's guidance supplied a written-permission default; RMIT directed students to course requirements without a fallback when courses remained silent. Macquarie adds another qualification: an AI-specific binding instrument supplied the operative rule even when a general binding instrument omitted GenAI, producing zero ambiguity.

## 4.6 Qualitative themes

Five overlapping themes recurred in the coded evidence (Table 5).

**Table 5. Qualitative themes and practice-oriented implications**

| Theme | Supporting rows | Practice-oriented implication |
|---|---|---|
| Permission-gated versus disclosure-based architectures | 37 | State the default architecture and whether the institution requires disclosure, treats it as sufficient for permission, or both. |
| Binding silence and guidance-dependent ambiguity | 100 | Publish a fallback rule and a clear hierarchy for institutional and task-level materials. |
| Retained text is clearer than process-only assistance | 80 | Add examples covering planning, source discovery, rewriting, and retained generated text. |
| Repurposed misconduct categories | 29 | Update guidance examples and cross-references without rewriting every durable misconduct category. |
| Verification failure and explicit task rules | 40 | Pair standard task-permission language with an explicit duty to verify claims and references. |

Note. Theme-support counts overlap and do not estimate mutually exclusive prevalence.

First, universities used either permission-gated or disclosure-based architectures. Permission-gated systems prohibited AI by default until a named authority or task instruction granted permission. Disclosure-based systems permitted some uses by default but imposed limits based on contribution or harm. The contrast helps explain Cases 4 and 7.

Institutions should state their default and clarify whether acknowledgment merely records use or also constitutes authorization.

Second, binding silence affected clarity only through the guidance that followed it. Guidance could supply a usable default, but open-ended delegation left students without an answer when educators had not spoken. Institutions should publish both a fallback rule and a hierarchy for resolving conflicts between institutional and task-specific material.

Third, policies regulated the retained GenAI text more clearly than process support did. Rules on authorship, editing, and original work often captured rewriting or generated paragraphs, but rarely clarified planning or verified source discovery. Guidance should include examples across that full continuum.

Fourth, existing misconduct categories often resolve GenAI cases. Contract cheating, ghostwriting, fabrication, and misrepresentation captured major drafting and fabricated references without new technology-specific rules. Institutions can update examples and cross-references while preserving the principles of durable integrity.

Fifth, verification duties and explicit assessment instructions cut across governance architectures. Standard assessment templates should combine clear AI-permission language with an express duty to verify claims, quotations, and references.

# 5. Discussion

## 5.1 From policy position to operational classification

The findings show why a single permissive-restrictive label cannot describe university GenAI governance. One institution may simultaneously encourage AI literacy, require educator permission, mandate disclosure, and apply older authorship or fabrication rules. Fixed cases reveal how those elements interact.

Policy layers, defaults, and operational clarity explain the pattern. Divergence may reflect deliberate choices among governance architectures. Indeterminacy instead signals an incomplete implementation chain: written materials delegate authority but fail to identify a default or fallback. This distinction links the results to regulatory intermediacy and the long-standing problem of rules and discretion. A system can therefore be restrictive and clear, permissive and clear, or ambiguous under either substantive orientation. The evaluative issue is not whether one default is ethically superior in every setting; it is whether students and educators can identify the default, the authority that may vary it, and the consequences of silence.

The structure matters more than the overall restrictive count. Universities unanimously honored an explicit prohibition on tasks, yet they diverged most on disclosed rewriting and a disclosed generated paragraph. They therefore handled local task authority more consistently than moderate levels of AI contribution.

The zero count for Clearly permitted follows from the codebook's strict definition. Many environments allowed GenAI only after acknowledgment, educator approval, or task authorization, so the analysis classified them as permitted with conditions. The result shows an absence of unconditional permission in these 300 case applications, not a sector-wide ban.

Most earlier studies evaluate content, institutional positions, or principles. Comparative policy-vignette testing adds an operational question: can the assembled written system support a defensible classification when conduct stays constant? It makes ambiguity observable rather than treating it as an impression of wording.

## 5.2 An operational classification model

The findings support a model in which written governance shapes decision conditions through two mechanisms (Figure 4). Policy layers allocate authority; defaults and fallbacks preserve continuity when lower-level instructions remain silent. Together they produce operational clarity, which structures whether students and educators proceed, disclose, seek approval, modify conduct, or refrain. This document study does not test the choices people actually make. The model also distinguishes policy content from policy performance. Content describes the principles or topics that documents mention. Performance concerns whether those materials, taken together, guide a reader from the facts of a case to a supportable answer. This second question makes defaults and handoffs visible.

**Fig. 4 Operational pathway from written governance to decision conditions**

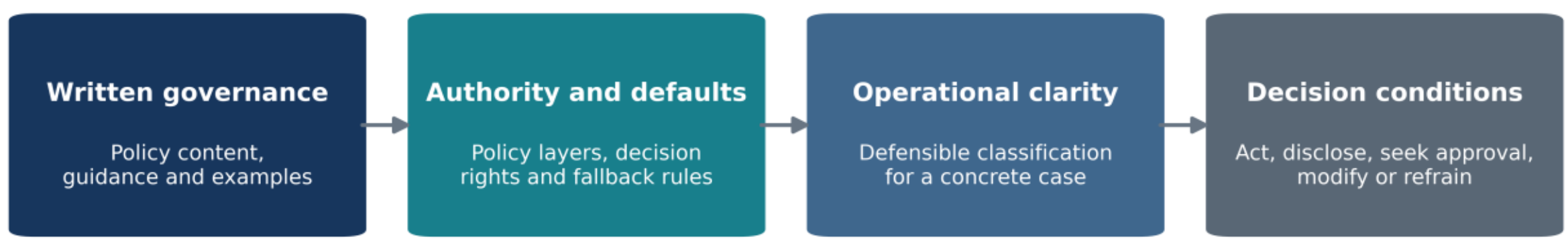

Note. Policy layers, defaults, and fallbacks shape operational clarity and decision conditions. This document study does not examine behavior.

The model shifts the unit of analysis from an isolated policy to a classification pathway through documents, authorities, and defaults. Conflicting layers weaken authority, missing local rules expose absent defaults, and vague contribution thresholds block classification even when authority remains clear. Different institutions may still reach different, equally clear outcomes because they chose different defaults. The model t, therefore, separates legitimate governance diversity from operational failure.

## 5.3 Competing governance architectures

Permission-gated systems treat silence as denial until a named authority grants approval. Disclosure-based systems treat transparency as the general condition unless a task rule says otherwise. Neither architecture always performs better. Permission gates can protect assessment security but create burdensome approval pathways; disclosure models can support AI literacy but may blur the boundary between acknowledgment and acceptable authorship.

These architectures explain why disclosure often changed little. Disclosure cannot cure unauthorized use in a permission-gated system, and it cannot cure prohibited substitution of generated work in a disclosure-based system. Institutions should state whether disclosure (a) acknowledges use, (b) forms a prerequisite for permission, (c) supplies evidence about authorship, or (d) serves more than one of these functions.

False declarations showed a similar ceiling effect: deception often aggravated conduct that already occupied a restrictive category. Institutions should separate the classification of a breach from later judgments about seriousness and sanction.

The two architectures also travel beyond Australia. Permission-gated systems need prompt, visible approval routes. Disclosure-based systems need clear thresholds for contribution and authorship. Both need an accessible hierarchy and a fallback rule so that students and educators can act when guidance or local instructions remain silent.

## 5.4 Guidance dependence and the problem of missing defaults

Guidance clarified most cases in which binding instruments remained silent, and technology-neutral binding rules can remain durable while guidance supplies timely examples. Risk arises when guidance conflicts across layers, claims unclear authority, or delegates decisions without a fallback. Effective guidance did four things: translated integrity principles into contribution-specific examples, stated a permission or disclosure default, named the role that could vary it, and pointed to identifiable task rules. Curtin and RMIT illustrate the effect. Curtin's short written-permission default produced little ambiguity; RMIT's open-ended delegation produced much more. Macquarie shows why researchers must review the entire binding environment rather than count GenAI keywords in a single general policy. Guidance also carries practical advantages: institutions can update examples more quickly than formal rules and tailor them to students or staff. Those advantages depend on version control, visible links to authority, and consistent wording across webpages. Without those safeguards, flexibility can create conflicting or unstable boundaries that users cannot reliably trace.

## 5.5 Retained text, verification, and assessment authority

The documents favored visible outputs. Policies often captured rewriting and retained generated paragraphs through rules on authorship, editing, or unauthorized assistance, but they less often addressed planning and verified source discovery. Institutions should decide explicitly which processes fall outside integrity restrictions, which require task-level approval, and which affect authorship.

Fabricated references produced more consistent outcomes because established duties of accuracy, verification, and non-fabrication already applied. Durable principles can govern new tools, but student guidance should still explain that plausible-looking AI citations require independent verification.

Case 15 demonstrates the force of assessment instructions: every university treated the explicit prohibition as controlling. This supports context-sensitive assessment but places a drafting burden on educators and can create within-institution variation. Standard templates, common permission categories, and centralized examples can preserve flexibility while reducing accidental inconsistency.

## 5.6 A practical policy decision framework

The results support a six-question drafting and audit framework (Table 6). Institutions may choose different substantive defaults, but their written environments should answer each question.

**Table 6. Operational framework for drafting and auditing GenAI assessment rules**

| Decision point | Question to answer | Minimum policy specification |
|---|---|---|
| 1. Task rule | Does the assessment explicitly permit, limit, or prohibit the use? | Provide a standard location and controlled vocabulary for task instructions. |
| 2. Default and authority | What applies if the task is silent, and who may vary that default? | State the fallback rule and name the approving role. |
| 3. Disclosure | Is acknowledgment required, and is it separate from permission? | Specify form, location, and consequences of missing or false disclosure. |
| 4. Contribution and authorship | Was AI used for process support, rewriting, retained content, or substantial drafting? | Give examples tied to learning outcomes and retained student responsibility. |
| 5. Verification | Were claims, quotations, and references independently checked? | State that students remain accountable for accuracy and non-existent sources. |
| 6. Hierarchy and escalation | Which instrument controls when policy, guidance, and task instructions differ? | Publish a conflict rule and an accessible route for timely clarification. |

The framework starts with task instructions because they produced the clearest result. It then asks for the institutional default and approval authority, separates disclosure from authorization, examines contribution and authorship, tests verification duties, and ends with hierarchy and escalation. Any unanswered question marks a predictable source of uncertainty.

At minimum, university materials should define use, assistance, generated content, editing, and authorization consistently; separate permission from acknowledgment; distinguish process support from retained content; name approval roles; require verification; and state which source controls when policy, guidance, and task instructions conflict. Worked vignettes can test whether readers can recover the intended answer.

Universities can use the framework to align academic-integrity guidance, GenAI strategy, and assessment templates. A workable rule set establishes the institutional default, provides educators with controlled options to vary it, standardizes disclosure language, and covers both process support and generated text.

Regulators can add policy-vignette stress tests to provider toolkits and assessment-reform resources. The test need not prescribe a single architecture; it asks whether the chosen model remains coherent and intelligible when task instructions are silent, in conflict, or expressly prohibit its use. A regulator could provide a small set of standard cases and ask providers to cite the document, authority, and fallback for each answer. Repeating the exercise after policy changes would reveal whether revisions improve clarity without requiring institutions to adopt identical substantive rules.

## 5.7 Contributions, limitations, and future research

Policy vignettes bridge content analysis and studies of actual decisions. By holding conduct constant, they show what written wording produces, distinguish formal authority from practical guidance, and expose ambiguity without forcing a judgment. The method can extend to AI-supported clinical documentation, legal drafting, research integrity, proprietary data use, and automated decision-making. The substantive categories will change, but the question remains: does the written system support a defensible classification? This study has five limitations. First, the purposive sample of 20 institutions does not statistically represent the Australian sector. Second, fixed vignettes simplify conduct and omit sanctions, evidentiary disputes, procedure, and local context. Third, the study captures public documents collected on July 21-22, 2026; policies may change, and internal materials may differ. Fourth, the independent human second coder achieved only moderate agreement with the human-led, AI-assisted primary coding, especially at the boundaries between Potential policy breach/Clearly prohibited and Indeterminate/Permitted with conditions. The analysis retains those disagreements as a reliability limitation and leaves the frozen primary dataset unchanged. Fifth, entropy measures category dispersion but cannot distinguish desirable diversity from problematic inconsistency. Future studies should double-code the full dataset, compare fully human coding with human-led AI-assisted coding, refine contested category thresholds, repeat collection over time, and test the framework with students, educators, and integrity decision-makers. Future research should also examine whether vignette-identified gaps predict difficulties observed in interviews, complaints, or actual integrity decisions; whether operational clarity improves comprehension and perceived fairness; and whether different disciplines consistently communicate educator discretion. Longitudinal replication could show whether institutions move toward clearer defaults as GenAI practices mature. Studies in other jurisdictions could test how regulatory traditions shape permission-gated and disclosure-based models without assuming that any one national approach serves as the benchmark.

## 6. Conclusion

Australian university GenAI governance does not follow a simple ban-permit divide. It combines binding rules, guidance, disclosure duties, established categories of misconduct, educator discretion, and assessment instructions. Across 20 universities, fixed vignettes produced many restrictive classifications and substantial uncertainty, but universities agreed when an assessment explicitly prohibited AI and generally applied existing accuracy rules to fabricated references.

Educator discretion works more predictably when institutions state a default, identify the decision maker, communicate task-level variation, and explain how permission relates to disclosure. Principle-based rules on authorship, accuracy, and fabrication remain useful, but guidance should illustrate both retained generated text and process-only assistance. Comparative policy vignettes do not predict disciplinary outcomes; they test whether a reasonable reader can recover a defensible classification from the written record. Clear and equitable governance, therefore, requires alignment among formal policy, official guidance, and assessment instructions.

## Data availability

Online Resource 1 contains the consolidated vignette instrument, codebook, coding protocol, and agreement summary. Online Resource 2 contains the anonymized 300-row dataset, an independent human second-coder sample, agreement details, derived results, and document evidence audit in a plain workbook. Online Resource 3 contains the analysis scripts and machine-readable inputs. Public university documents remain subject to their source institutions' terms and to later revision.

# Statements and Declarations

Competing interests: The author declares no competing interests.

Funding: No funding supported this study.

Ethics approval: Not applicable. The study used public institutional documents and involved no human participants or personal data.